\overfullrule=0pt
\input harvmac

\Title{\vbox{\hbox{IFT-P.046/99}}}
{\vbox{
\centerline{\bf On the N=2 Superstring BRST Operator}}}

\bigskip\centerline{Osvaldo Chand\'{\i}a\foot{e-mail:
chandia@ift.unesp.br}}
\bigskip
\centerline{\it Instituto de F\'\i sica Te\'orica, Universidade Estadual
Paulista}
\centerline{\it Rua Pamplona 145, 01405-900, S\~ao Paulo, SP, Brasil}

\vskip .3in
We show that the BRST charge for the N=2 superstring system can be
written as $Q=e^{-R} (\oint{dz \over 2\pi i} b\gamma_+\gamma_-)e^R $,
where $b$ and $\gamma_\pm$ are super-reparametrizations ghosts. This provides
a trivial proof of the nilpotence of this operator.

\Date {May, 1999}

\newsec{Introduction}

Superstring theory can be seen as a critical $N=1$ superconformal system
defined on the world-sheet. It can be quantized by studying the
cohomology of the nilpotent BRST operator
\eqn\qone{ Q=\oint {dz \over 2\pi i}[c(T+\half T^g)+\gamma (G+\half G^g)]
,}
where the matter generators $T$ and $G$ satisfy the $N=1$ superconformal
algebra with central charge $c=15$, and the ghosts generators $T^g$ and
$G^g$ satisfy the corresponding algebra with central charge $c=-15$. In
such a way, the ghost sector allows to fix the gauge symmetry of the
theory. The ghosts generators are constructed out of a pair of fermionic
fields $b, c$ (with spins $2$ and $-1$, respectively) that fix the
conformal symmetry, and a pair of
bosonic fields $\beta, \gamma$ (with spins $3/2$ and $1/2$, 
respectively) that fix the world-sheet supersymmetry.

Physical states are described by vertex operators living in the
cohomology of the BRST operator \qone. In order to include spacetime
spinors in the spectrum, we need to fermionize the bosonic ghosts as
\eqn\ferm{\beta = \partial \xi e^{-\phi},\quad \gamma = \eta e^{\phi},}
where $\eta$ and $\xi$ are free fermions of spins $1$ and $0$
respectively, and $\phi$ is a chiral boson \ref\fms{D. Friedan, E.
Martinec and S. Shenker, ``Conformal Invariance, Supersymmetry and
String Theory,'' Nucl. Phys. {\bf B271}, 93 (1986).}. Note that this
fermionization does not involve the zero mode of $\xi$, then physical
states are independent of this mode (it is due to the fact that physical
states are constructed out of $\beta$ and not out of $\xi$, for example).
The space of the physical states is called ``small'' Hilbert space and
the whole space, involving operators constructed out of $\xi$ zero mode, 
is called ``large'' Hilbert space. We can define the space of physical
states as the set of operators that commute with $\oint \eta$. Then,
physical vertex operators not only must belong to the cohomology of $Q$
but also must commute with $\oint \eta$. As consistency, $Q$ not only
must be nilpotent but also must anticommute with $\oint \eta$
\ref\berkovits{N. Berkovits, ``A New Description of the Superstring,''
Jorge Swieca Summer School 1995, p.490. hep-th/9694123.}.

A similar analysis can be carried out for the string with N=2
world-sheet superconformal symmetry \ref\gr{A. Giveon and M. Rocek, ``On 
the BRST Operator Structure of the N=2 String,'' Nucl. Phys. {\bf
B400}, 145 (1993).}. Physical states belong to the cohomology of the
N=2 BRST operator 
\eqn\two{ Q=\oint {dz \over 2\pi i}[c(T+\half T^g)+\gamma_+ (G_- +\half
G_-^g)+\gamma_+(G_+ +\half G_+^g)+{\tilde c}(J+\half J^g)] ,}
where the matter generators $T$, $J$ and $G_\pm$ satisfy the N=2
superconformal algebra with central charge $c=6$ and the ghosts
generators $T^g$, $J^g$ and $G^g_\pm$ satisfy the corresponding algebra
with central charge $c=-6$. The critical system can be represented by a
pair of complex chiral superfields ${\bf X}^i({\bar z}, z, \theta^+,
\theta^-)$ and
${\bf  \bar X}^i(z, {\bar z}, {\bar\theta}^+, {\bar \theta}^-)$ ($i=1,2$),
being the signature of the background space ($2,2$) or Euclidean, but
not Minkowskian. This theory describes self-dual systems in four
dimensions (for related issues see \ref\ov{H. Ooguri and C. Vafa,
``Geometry of N=2 Strings,'' Nucl. Phys. {\bf B361}, 469 (1991).}). The
ghosts generators are constructed out of a pair of fermionic ghosts $b$,
$c$
(with spins $2$, $-1$, respectively) that fix conformal symmetry,
and another
pair of fermionic ghosts ${\tilde b}$, ${\tilde c}$ (with spins $1$ and $0$,
respectively) that fix the $U(1)$ gauge symmetry generated by $J$, and
four bosonic ghosts $\beta_\pm$ and $\gamma_\pm$ (with spins $3/2$ and
$-1/2$, respectively) that fix the world-sheet supersymmetries. 

We need to fermionize the bosonic ghosts in order to describe spacetime
spinors, the pair $\beta_+$ and $\gamma_-$ becomes
\eqn\fermtwo{\beta_+ = \partial \xi_+ e^{-\phi_+} , \quad \gamma_- =
\eta_+ e^{\phi_+},}
where $\xi_+$ and $\eta_+$ (spins $0$ and $1$, respectively) are free
fermions, and $\phi_+$ is a chiral boson. For the
bosonic pair $\beta_-$ and $\gamma_+$ there is an equivalent expression
involving $\xi_-$, $\eta_-$ and $\phi_-$ instead $\xi_+$, $\eta_+$ and
$\phi_+$. Note that, as in the N=1 case, the zero modes of the ghosts
$\xi_\pm$ are not involved in the fermionizations, then physical states
are independent of such modes. We can define the space of physical states
as the set of vertex operators that commute with $\oint \eta_+$ and
$\oint \eta_-$. This will be the analogous of the ``small'' Hilbert space
of the N=1 case, the ``large'' Hilbert space takes into account
operators that depend on zero modes of $\xi_\pm$. As consistency, the
BRST operator must be not only nilpotent but also anticommute with
$\oint \eta_\pm$.

It was shown in \ref\abc{J. Acosta, N. Berkovits and O. Chand\'{\i}a,
``A Note on the
Superstring BRST Operator,'' Phys. Lett. {\bf B} (to appear).
hep-th/9902178.} that the N=1 superstring BRST operator can be written as 
\eqn\qb{Q=e^{-R}(\oint {dz \over 2\pi i} b\gamma^2)e^R,}
which trivially proves the nilpotence of the BRST operator. This also
shows that the cohomologies of $Q$ and $\oint {dz \over 2\pi i}
b\gamma^2$ are equals. The last one is trivial in the ``large'' Hilbert
space, then $Q$ is trivial in this space. In the ``small'' $Q$ is not
trivial as expected.

The purpose of this paper is to extend the result \qb\ for the N=2
superstring.

\newsec{Similarity Transformation for the N=2 Superstring} 

The BRST current $j_{BRST}(z)$ is given by $Q=\oint {dz \over 2\pi
i}j_{BRST}(z)$. After fermionize the bosonic ghosts as in \fermtwo\ and
then bosonize $\xi_\pm = e^{\chi_\pm}$ and $\eta_\pm = e^{-\chi_\pm}$, 
the
BRST current becomes \gr\
\eqn\jtwo{\eqalign{&j_{BRST} = cT + e^{\phi_+ - \chi_+}G_- + e^{\phi_-
-\chi_-}G_+ + {\tilde c} J + bc\partial c - c {\tilde b}\partial{\tilde c} \cr
& \half {\tilde b} [e^{\phi_+ - \chi_+}\partial(e^{\phi_- - \chi_-}) -
e^{\phi_- - \chi_-}\partial(e^{\phi_+ - \chi_+})] - 2 b e^{\phi_- - \chi_-}
e^{\phi_+ - \chi_+} \cr
& + c T^{\phi \chi} + {\tilde c} (\partial \phi_+ -
\partial \phi_-) ,\cr}}
where 
$$T^{\phi \chi}=-{\partial}^2\phi_+ - \half (\partial\phi_+)^2 + \half
{\partial}^2\chi_+ + \half (\partial\chi_+)^2$$
$$ -{\partial}^2\phi_- - \half (\partial\phi_-)^2 + \half 
{\partial}^2\chi_- + \half (\partial\chi_-)^2.$$
The BRST current has a total derivative term that we have no written in
\jtwo\ .

We will show that 
\eqn\simil{j_{BRST} = e^{-R} j_0 e^R ,}
where
\eqn\nzero{j_0 = -2b e^{\phi_- - \chi_-} e^{\phi_+ - \chi_+},}
and
$$R = \half \oint {dz \over 2\pi i} [cG_+e^{-\phi_-+\chi_-} +   
cG_-e^{-\phi_++\chi_+}]$$
$$-\half \oint {dz \over 2\pi i} c\partial{\tilde c} (H +\phi_+ -
\phi_-)e^{-\phi_++\chi_+}e^{-\phi_-+\chi_-}$$ 
$$-{1\over 4}\oint {dz \over 2\pi i} c\partial c
[\partial(e^{-\phi_-})e^{\chi_-}e^{-\phi_+ + \chi_+} +  
\partial(e^{-\phi_+})e^{\chi_+}e^{-\phi_- + \chi_-}]$$
$$+\half\oint {dz \over 2\pi i} c{\tilde b} [e^{-\phi_- + \chi_-}\partial
(e^{\phi_- - \chi_-}) - e^{-\phi_+ + \chi_+}\partial (e^{\phi_+ - \chi_+})],$$
where the $U(1)$ current is bosonized as $J \equiv \partial H$. Using
\simil, $j_{BRST}$ is trivially nilpotent since $j_0$ has no poles with
itself.

To prove \simil\ we use the expansion
\eqn\expa{e^{-R}j_0e^R \equiv \sum_{n=0}^{\infty} {1 \over n!}j_n,}
$$j_n = [j_{n-1} , R],$$
where, for $R = \oint {dz \over 2\pi i}r(z)$ the commutator is computed
using the rule
$$[j_{n-1}(y) , R] = \oint {dz \over 2\pi i} j_{n-1}(y) r(z).$$

The term  $n=1$ in \expa\ is given by

\eqn\none{\eqalign{&j_1=- {3 \over 2}{\partial}^2 c + e^{\phi_+-\chi_+}G_+ +
e^{\phi_--\chi_-}G_- + bc\partial c - \partial{\tilde c} (H +\phi_+ - \phi_-)
\cr    & + \partial c (3\partial \phi_+ + 3\partial \phi_- -
2\partial \chi_+ - 2\partial \chi_-)+{\tilde b} [e^{\phi_+ - \chi_+}\partial
(e^{\phi_- - \chi_-}) - e^{\phi_- - \chi_-}\partial(e^{\phi_+ - \chi_+})]\cr   
& + c({\partial}^2 \phi_+ + {\partial}^2 \phi_-
- \half {\partial}^2 \chi_+ - \half {\partial}^2 \chi_- - (\partial \phi_+)^2
-(\partial \phi_-)^2 - \half (\partial \chi_+)^2 - \half (\partial \chi_-)^2
\cr  
& - 2 \partial \phi_+ \partial \phi_- + {3\over 2}\partial \phi_+\partial
\chi_+ + {3\over 2}\partial \phi_+\partial \chi_- + {3\over 2}\partial \phi_-\partial\chi_+ +
{3\over 2}\partial\phi_-\partial\chi_- - \partial\chi_+\partial\chi_-),\cr}}  
the term $n=2$ is given by 
\eqn\ntwo{\eqalign{&j_2= \half {\partial}^2c + 2cT -2c{\tilde
b}\partial{\tilde c} - \half G_+  c\partial c e^{-\phi_- + \chi_-} -
\half G_- c\partial c e^{-\phi_+ + \chi_+} + c\partial c
\partial^2 c e^{-\phi_+ + \chi_+}e^{-\phi_- + \chi_-} \cr
& + {\tilde b} c\partial c (\partial\phi_+ -
\partial\phi_- - \partial\chi_+ + \partial\chi_-) + 2\partial c
(-\partial\phi_+ - \partial\phi_- + \partial\chi_+ + \partial\chi_-) \cr 
& + c( (\partial \phi_+)^2  + (\partial \phi_-)^2 + 2 (\partial \chi_+)^2 +
2 (\partial \chi_-)^2 + 4 \partial \phi_+\partial \phi_- - 3 \partial
\phi_+\partial \chi_+ - 3 \partial \phi_+\partial \chi_- \cr
& - 3 \partial
\phi_-\partial\chi_+ - 3 \partial\phi_-\partial\chi_- +
2 \partial\chi_+\partial\chi_+)  
+ \half \partial{\tilde c} c\partial c
(H + \phi_+ - \phi_-) e^{-\phi_+ + \chi_+}e^{-\phi_- + \chi_-} ,\cr}}
the term $n=3$ is given by 
\eqn\nthree{\eqalign{&j_3={3\over 2}G_+c\partial c
e^{-\phi_- + \chi_-} + {3\over 2}G_-c\partial c e^{-\phi_+ + \chi_+} + 3
{\tilde b} c\partial c (-\partial\phi_+ + \partial\phi_- + \partial\chi_+ -
\partial\chi_-) \cr 
& -{3\over 4} c\partial c {\partial}^2 c e^{-\phi_+ +
\chi_+}e^{-\phi_- + \chi_-} -{3\over 2} \partial{\tilde c} c\partial c
(H + \phi_+ - \phi_-) e^{-\phi_+ + \chi_+}e^{-\phi_- + \chi_-} ,\cr}}
and the term $n=4$
\eqn\nfour{j_4=-15 c\partial c {\partial}^2 c e^{-\phi_+ +
\chi_+}e^{-\phi_- + \chi_-}.}
The term for $n=5$ in the expansion vanishes identically since the OPE between
$j_4$ and $R$ has no single poles. Then, the terms of higher order in the
expansion \expa\ vanish too.

It is straightforward to check that the BRST current is equal to $j_0 + j_1 +
{1\over 2!}j_2 + {1\over 3!}j_3 +{1\over 4!}j_4$ up to total derivatives. 

\newsec{Concluding Remarks}

The form that we have written the BRST current \simil, proves its
nilpotence trivially. Note that the cohomology of $\oint j_0$ is trivial
in the ``large'' Hilbert space, then the N=2 cohomology is trivial in
that
space. However, such a property is not hold in the ``small'' Hilbert
space since the N=2 has non-trivial states \ov. 

One could be tempted to use the expansion \expa\ as BRST current for the
non-critical case. However, if the central charge $c$ is different of the
critical value $6$, in the expansion would appear one term proportional to
$(6-c)\oint dz c\partial c \partial^2 c e^{-\phi_+ + \chi_+} e^{-\phi_- +
\chi_-}$ which does not commute with $\oint dz \eta_\pm = \oint dz
e^{\chi_\pm}$. Therefore, the expansion \expa\ can be used as BRST charge in
the critical case only. 

\vskip 10pt
{\bf Acknowledgements:} I would like to thank Nathan
Berkovits for useful comments and suggestions. This work was supported by
FAPESP grant 98/02380-3.

\listrefs

\end